\documentclass[conference]{IEEEtran}
\IEEEoverridecommandlockouts

\usepackage{cite}
\usepackage{amsmath,amssymb,amsfonts}
\usepackage{algorithmic}
\usepackage{algorithm}
\usepackage{balance}
\usepackage{bm}
\usepackage{booktabs}
\usepackage[table]{xcolor}
\definecolor{Best}{RGB}{232,222,248}
\usepackage{graphicx}
\usepackage{textcomp}
\usepackage{xcolor}
\def\BibTeX{{\rm B\kern-.05em{\sc i\kern-.025em b}\kern-.08em
    T\kern-.1667em\lower.7ex\hbox{E}\kern-.125emX}}
\begin{document}

\title{GeoFovea-GS: Geometry-Aware Cross-Layer Gaussian Splatting for Wireless Aerial VR
\thanks{This work is supported in part by the State Key Laboratory of Internet of Things for Smart City (University of Macau) Open Research Project under Grant SKL-IoTSC(UM)/ORP04/2026, and in part by the Project of Tsinghua University-Toyota Joint Research Center for AI Technology of Automated Vehicle under Grant TTAD-2024-08-2.}
}

\author{
   \IEEEauthorblockN{
Zeyi Ren,
Wencheng Yan,
Jiawen Zhang,
Jintao Yan,\\
Sheng Zhou, \IEEEmembership{Senior Member, IEEE},
and Zhisheng Niu, \IEEEmembership{Fellow, IEEE}
}
    
    \IEEEauthorblockA{Department of Electronic Engineering, Tsinghua University, Beijing, China}


%
    \IEEEauthorblockA{Emails: zeyiren0827@outlook.com,  \{sheng.zhou@, niuzhs@\}tsinghua.edu.cn}
}

\maketitle

\begin{abstract}
Wireless aerial virtual reality (VR) aims to provide immersive access to large-scale scenes, but high-resolution view generation and delivery are jointly constrained by limited bandwidth, latency, and power. 3D Gaussian Splatting (3DGS) can reduce the payload by rendering views from compact pose information, yet its geometry errors may cause severe VR quality degradation. Existing channel-aware or pixel-level resource allocation schemes fail to capture such geometry-sensitive distortion. To address this issue, this paper proposes GeoFovea-GS as a geometry-aware cross-layer framework for communication-efficient wireless aerial VR. A foveated geometry-aware distortion metric is developed to characterize photometric rendering error, geometric inconsistency, and view-dependent perceptual importance in a unified form. Based on this metric, the joint selection of pose-only 3DGS rendering and image/tile correction transmission is formulated as a cross-layer optimization problem under wireless constraints. A lightweight value-of-information scheduler is further developed to allocate communication resources to regions that are both geometry-critical and perceptually important. Experiments on real-world 3DGS scenes demonstrate that GeoFovea-GS achieves superior immersive rendering quality with substantially reduced transmission cost.
\end{abstract}


\section{Introduction}

Wireless aerial virtual reality (VR) is becoming an important technology for remote exploration~\cite{wang2025multirobotcooperativeexplorationunknown}, digital twins~\cite{jiawen_tgpp}, infrastructure inspection~\cite{icra_inspection}, and low-altitude intelligent networks~\cite{gaofei,LAIN_zhou}. By reconstructing large-scale scenes captured by unmanned aerial vehicles (UAVs), users can interactively observe outdoor environments from flexible six-degree-of-freedom viewpoints. However, practical aerial VR systems require high-resolution transmission, low latency, and stable visual quality, which impose severe burdens on wireless bandwidth and transmit power~\cite{ren2026efficienttransceiver}.

\begin{figure} [t]
	\centering 
		 \includegraphics[width=0.48\textwidth]{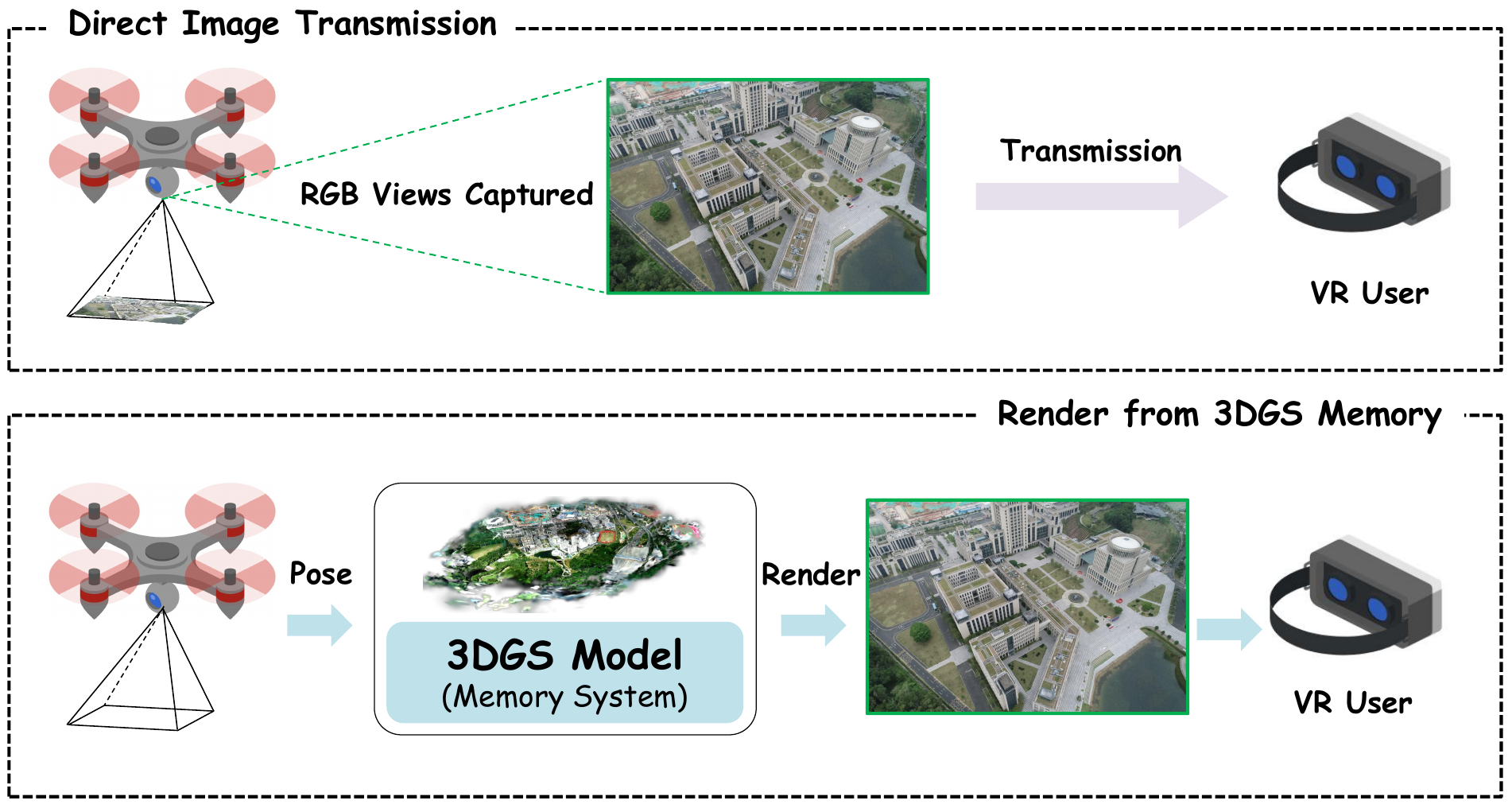} 
  	\caption{Conceptual illustration of wireless aerial VR. A UAV captures large-scale outdoor scenes, which are reconstructed into a digital environment as a memory system and delivered over wireless links for immersive remote exploration.}\vspace{-0.2in}  \label{Fig1}
\end{figure}

Recent advances in 3D Gaussian Splatting (3DGS) provide a new opportunity to reduce this communication burden~\cite{kerbl3Dgaussians,STT-GS}. Instead of transmitting every image frame, the system can transmit compact pose information and synthesize the corresponding view from a pre-trained 3DGS scene representation. This memory-assisted paradigm changes the transmitted content from image-level data to pose-level data, making it attractive for bandwidth-limited wireless VR systems~\cite{shuaiXR,chen2018vr}.

Nevertheless, directly relying on 3DGS rendering is unreliable in large-scale aerial scenes~\cite{wang2026lagslowaltitudegaussiansplatting}. UAV trajectories are often sparse and irregular, while outdoor scenes contain long-range structures and depth discontinuities. As a result, 3DGS may generate views with plausible appearance but incorrect geometry. These geometry errors can be damaging to VR experience, since they cause view inconsistency and incorrect spatial cues during user interaction.

Existing wireless resource allocation methods are insufficient for geometry-sensitive aerial VR. Classical channel-aware schedulers allocate radio resources according to channel gains or achievable rates~\cite{yc-allocation}, while wireless VR resource allocation further optimizes latency~\cite{yichen-globecom}, Quality of Experience (QoE), or viewport utility~\cite{chen2023crossframe,feng2023qoe}. However, these objectives do not explicitly model the reliability of the reconstructed 3D scene. Image quality-aware schemes commonly rely on viewport-level peak signal-to-noise ratio (PSNR), structural similarity index (SSIM), or rate-distortion utilities~\cite{chakareski2020viewport}, although such pixel-level metrics can disagree with human perceptual judgments. Foveated and viewport-adaptive streaming reduces bandwidth by prioritizing the predicted field of view (FoV) or gaze region~\cite{li2021log}, but it still treats visual content as video tiles rather than uncertain renderings from a 3D representation. In contrast, recent neural rendering studies show that sparse or unbounded views may lead to degenerate geometry, depth artifacts, or scale-dependent rendering artifacts~\cite{niemeyer2022regnerf,wang2023sparsenerf,yu2024mipsplatting}, which motivates geometry-aware wireless correction.

To overcome these limitations, this paper proposes GeoFovea-GS, a
geometry-aware cross-layer framework for wireless aerial VR with 3DGS. The
central idea is to spend wireless resources only on regions where geometry
unreliability, perceptual importance, and channel affordability coincide.
The main contributions are summarized as follows.
\begin{itemize}
    \item A foveated geometry-aware distortion metric for 3DGS-assisted aerial VR is developed. The metric jointly captures photometric rendering error,
    geometry inconsistency, temporal instability, and view-dependent perceptual
    importance, providing a tile-level reliability map for wireless correction.

    \item This paper formulates the joint selection of pose-only 3DGS rendering and
    image/tile correction transmission as a cross-layer optimization problem
    under rate, latency, average-power, and peak-power constraints. By deriving
    the minimum power required for a fixed tile selection, the original
    mixed-integer problem is transformed into a content-selection problem with
    channel-dependent power costs.

    \item We design a lightweight value-of-information scheduler that allocates
    wireless resources to tiles with large geometry-aware perceptual benefit
    and low incremental power cost. Experiments on real-world UAV scenes show
    that GeoFovea-GS improves immersive rendering quality and reduces
    communication cost compared with GS-only rendering, full image
    transmission, and conventional resource allocation baselines.
\end{itemize}

\section{System Model}

A wireless aerial VR system is considered, where a server provides immersive views of a large-scale UAV-captured scene to a VR user. The scene is represented by a pre-trained 3D Gaussian Splatting (3DGS) model. The system operates over $T$ time slots, indexed by $\mathcal{T}=\{1,\ldots,T\}$. At time slot $t$, the requested VR viewpoint is denoted by
\begin{equation}
    \mathbf{s}_t = [\mathbf{q}_t^T,\bm{\theta}_t^T]^T,
\end{equation}
where $\mathbf{q}_t\in\mathbb{R}^3$ denotes the camera position and $\bm{\theta}_t\in\mathbb{R}^3$ denotes the camera orientation.

\subsection{3DGS-Assisted Aerial VR Rendering}

Let $\Phi$ denote the pre-trained 3DGS model. Given viewpoint $\mathbf{s}_t$, the 3DGS-rendered view is
\begin{equation}
    \mathbf{y}_t = \Phi(\mathbf{s}_t),
\end{equation}
where $\mathbf{y}_t\in\mathbb{R}^{H\times W\times 3}$ is the synthesized RGB image with height $H$ and width $W$. The corresponding reference image is denoted by $\mathbf{r}_t\in\mathbb{R}^{H\times W\times 3}$.

To support fine-grained wireless content adaptation, each image is divided into $K$ non-overlapping tiles, indexed by $\mathcal{K}=\{1,\ldots,K\}$. Let $\mathbf{y}_{t,k}$ and $\mathbf{r}_{t,k}$ denote the $k$-th tile of the rendered image and the reference image, respectively.

At each tile, the system has two choices. If the 3DGS rendering is trusted, only the compact viewpoint information is transmitted. If the 3DGS rendering is unreliable, an image correction is transmitted for this tile. The binary content selection variable is defined as
\begin{equation}
    z_{t,k} =
    \begin{cases}
        1, & \text{if tile } k \text{ at slot } t \text{ is transmitted},\\
        0, & \text{if tile } k \text{ is rendered by 3DGS}.
    \end{cases}
\end{equation}

When tile $k$ at time slot $t$ is selected for transmission, the server sends a correction tile $\tilde{\mathbf r}_{t,k}$ instead of relying on the 3DGS-rendered tile $\mathbf y_{t,k}$. The correction tile represents the best available
scene-side reconstruction for the requested view. In the experiments, the
held-out reference tile is adopted as an oracle high-quality correction, which
provides an upper bound on correction performance. The displayed VR tile is
therefore
\begin{equation}
\mathbf m_{t,k}
=
z_{t,k}\tilde{\mathbf r}_{t,k}
+
(1-z_{t,k})\mathbf y_{t,k}.
\end{equation}
Thus, $z_{t,k}=1$ replaces the unreliable 3DGS-rendered tile with a correction tile, while $z_{t,k}=0$ saves communication cost by relying on pose-only 3DGS
rendering.

\subsection{Foveated Geometry-Aware Distortion}

For aerial VR, pose-only 3DGS rendering may induce photometric errors, geometric inconsistency, and temporal instability. The resulting 3DGS-induced distortion of tile $k$ at time slot $t$ is defined as
\begin{equation}
d^{\mathrm{GS}}_{t,k}
=
\alpha \bar d^{p}_{t,k}
+
\beta \bar d^{g}_{t,k}
+
\gamma \bar d^{\mathrm{temp}}_{t,k},
\end{equation}
where $\bar d^{p}_{t,k}$, $\bar d^{g}_{t,k}$, and
$\bar d^{\mathrm{temp}}_{t,k}$ are the normalized photometric distortion,
geometry inconsistency, and temporal instability, respectively. The
non-negative weights $\alpha$, $\beta$, and $\gamma$ control their relative
importance. 

Since the three raw distortion terms have different numerical scales, each term is normalized as
\begin{equation}
\bar d^{x}_{t,k}
=
\frac{d^{x,\mathrm{raw}}_{t,k}}{\mu_x+\varepsilon_x},
\quad
x\in\{p,g,\mathrm{temp}\},
\label{eq:normalization}
\end{equation}
where $\mu_x$ is the average value of the corresponding raw distortion over the
scheduling window, and $\varepsilon_x$ is a small constant for numerical
stability.

Unless otherwise specified, we use equal weights
$\alpha=\beta=\gamma=1$ after the normalization in \eqref{eq:normalization}.
This setting avoids manual scale tuning, because each raw term is divided by
its average value over the scheduling window. In practice, moderate changes of
these weights mainly adjust the priority among photometric, geometric, and
temporal corrections, while the final selected tiles are still jointly governed
by the foveated weight and the channel-dependent power cost.

The raw photometric distortion is computed as
\begin{equation}
d^{p,\mathrm{raw}}_{t,k}
=
\frac{1}{|\Omega_k|}
\sum_{\mathbf u\in\Omega_k}
\left\|
\mathbf x_t(\mathbf u)-\mathbf y_t(\mathbf u)
\right\|_2^2,
\label{eq:photo_dist}
\end{equation}
where $\Omega_k$ is the pixel set of tile $k$, $\mathbf u$ denotes a pixel
coordinate, and $\mathbf x_t$ denotes the correction/reference image available
for distortion estimation. In offline evaluation, $\mathbf x_t$ is set to the
held-out reference image $\mathbf r_t$.

\begin{figure*} [t]
	\centering 
		 \includegraphics[width=0.98\textwidth]{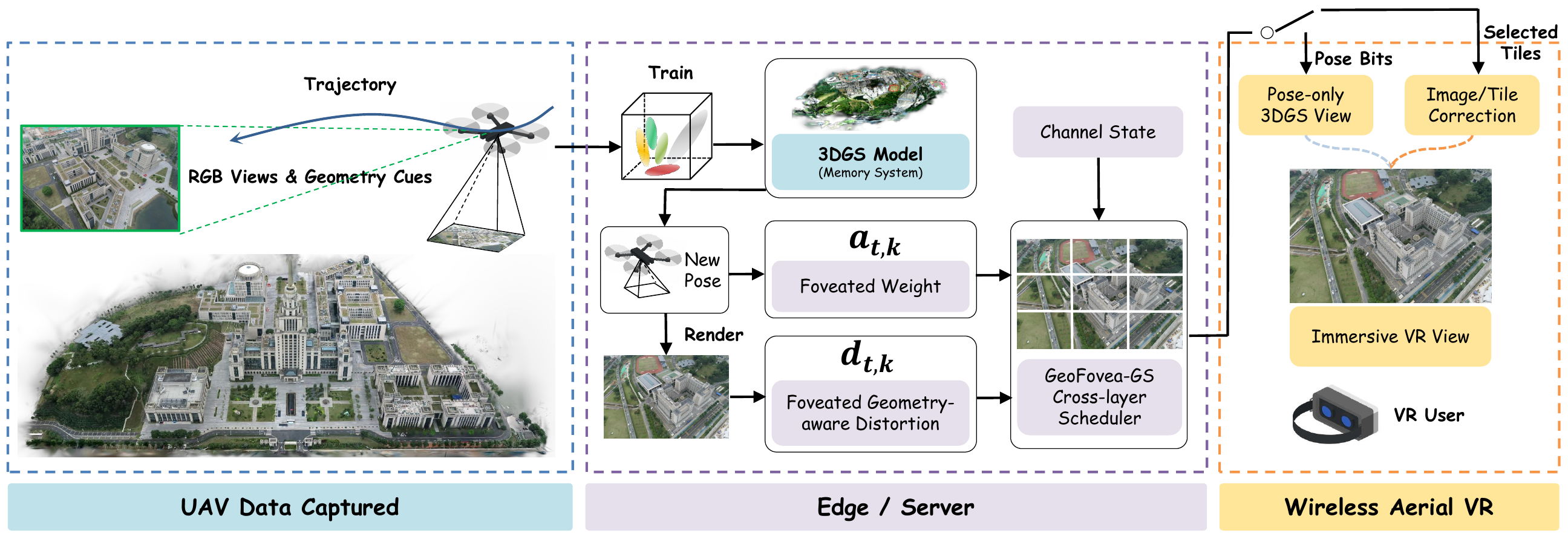}  
	
 	\caption{Workflow of the proposed GeoFovea-GS framework. The system uses a pre-trained 3DGS model as scene memory for pose-only rendering, estimates foveated geometry-aware distortion for each view or tile, and allocates wireless resources to transmit corrections only for perceptually important regions.}  \label{Fig2}
\end{figure*}

Let $Z^{\mathrm{ref}}_t(\mathbf u)$ denote the reference geometry cue projected
onto the query image plane, and let $\hat Z_t(\mathbf u)$ denote the depth
rendered from 3DGS. The valid geometry set of tile $k$ is
\begin{equation}
\Omega^g_{t,k}
=
\left\{
\mathbf u\in\Omega_k:
Z^{\mathrm{ref}}_t(\mathbf u)>0,\;
\hat Z_t(\mathbf u)>0
\right\}.
\end{equation}
If the geometry cue is only available up to scale, the rendered depth is aligned by
\begin{equation}
\lambda_t
=
\operatorname{median}_{\mathbf u\in\Omega^g_t}
\frac{Z^{\mathrm{ref}}_t(\mathbf u)}
{\hat Z_t(\mathbf u)+\varepsilon_z}.
\end{equation}
For metric depth cues, we set $\lambda_t=1$. The raw geometry inconsistency is
defined as
\begin{equation}
\begin{aligned}
d^{g,\mathrm{raw}}_{t,k}
&=
\frac{1}{|\Omega^g_{t,k}|}
\sum_{\mathbf u\in\Omega^g_{t,k}}
\min\!\left(\delta^g_t(\mathbf u),\rho_{\max}\right), \\
\delta^g_t(\mathbf u)
&=
\left|
\log\!\left(Z^{\mathrm{ref}}_t(\mathbf u)+\varepsilon_z\right)
-
\log\!\left(\lambda_t\hat Z_t(\mathbf u)+\varepsilon_z\right)
\right|.
\label{eq:geo_dist}
\end{aligned}
\end{equation}
For tiles without valid geometry pixels, the geometry term is ignored and the
scheduler relies on the photometric and temporal terms.

The temporal term is used as a scheduling proxy that identifies tiles whose
pose-only 3DGS rendering is temporally unstable. It is computed as
\begin{equation}
d^{\mathrm{temp},\mathrm{raw}}_{t,k}
=
\frac{1}{|\Omega_k|}
\sum_{\mathbf u\in\Omega_k}
\left\|
\mathbf y_t(\mathbf u)
-
\mathcal W_{t-1\rightarrow t}
\left(\mathbf y_{t-1}\right)(\mathbf u)
\right\|_2^2,
\end{equation}
where $\mathcal W_{t-1\rightarrow t}(\cdot)$ denotes the warping operation from
viewpoint $s_{t-1}$ to $s_t$. This term prioritizes tiles whose 3DGS rendering
is likely to cause unstable VR perception.

The three distortion terms have different information requirements. During
online scheduling, $d^{p,\mathrm{raw}}_{t,k}$ is estimated from the available
correction candidate, $d^{g,\mathrm{raw}}_{t,k}$ is computed from scene-side
geometry cues such as pre-computed depth, SfM/LiDAR projections, or 3DGS
uncertainty, and $d^{\mathrm{temp},\mathrm{raw}}_{t,k}$ is obtained directly
from consecutive 3DGS-rendered views and the known pose change. In offline
evaluation, we set $\mathbf{x}_t=\mathbf{r}_t$ and use the available scene depth
as $Z_t^{\mathrm{ref}}$, so the reported results represent an oracle upper
bound on the benefit of geometry-aware wireless correction.

Since VR users are more sensitive to the center of gaze, a foveated weight $a_{t,k}\in[0,1]$ is assigned to tile $k$. A common choice is
\begin{equation}
    a_{t,k}
    =
    \exp
    \left(
    -\frac{\|\mathbf{c}_{k}-\mathbf{g}_{t}\|_2^2}{2\sigma_f^2}
    \right),
\end{equation}
where $\mathbf{c}_{k}$ is the center coordinate of tile $k$, $\mathbf{g}_t$ is the gaze center, and $\sigma_f$ controls the foveation range.

Let $d^{\mathrm{corr}}_{t,k}$ denote the residual distortion after transmitting
the correction tile. For high-quality correction transmission,
$d^{\mathrm{corr}}_{t,k}$ is small and can be set to zero. The remaining VR
distortion at time slot $t$ is
\begin{equation}
D_t(\mathbf z_t)
=
\sum_{k=1}^{K}
a_{t,k}
\left[
(1-z_{t,k})d^{\mathrm{GS}}_{t,k}
+
z_{t,k}d^{\mathrm{corr}}_{t,k}
\right].
\end{equation}

\subsection{Wireless Transmission Model}

Let $C^{\mathrm{pose}}$ denote the number of bits required to transmit the viewpoint information, and let $C_{t,k}$ denote the number of bits required to transmit tile $k$ at time slot $t$. The total transmitted data size at slot $t$ is
\begin{equation}
    C_t(\mathbf{z}_t)
    =
    C^{\mathrm{pose}}
    +
    \sum_{k=1}^{K} z_{t,k} C_{t,k}.
\end{equation}

We consider an uplink or downlink wireless channel with bandwidth $B$. Let $h_t\in\mathbb{C}$ denote the channel coefficient at time slot $t$, and let $p_t\geq0$ denote the transmit power. The achievable rate is
\begin{equation}
    R_t(p_t)
    =
    B\log_2
    \left(
    1+\frac{|h_t|^2p_t}{\sigma^2}
    \right),
\end{equation}
where $\sigma^2$ is the noise power. To successfully deliver the selected content within slot duration $\tau$, the following constraint should hold:
\begin{equation}
    \tau R_t(p_t) \geq C_t(\mathbf{z}_t), \quad \forall t\in\mathcal{T}.
\end{equation}

\section{Problem Formulation and Optimization Algorithm}

\subsection{Cross-Layer Optimization Problem}

The objective is to minimize the average foveated geometry-aware VR distortion under wireless power and latency constraints. The optimization variables are the tile selection variables $\{z_{t,k}\}$ and transmit powers $\{p_t\}$. The problem is formulated as
\begin{subequations}
\begin{align}
\mathbf{P0}: \quad
\min_{\{z_{t,k}\},\{p_t\}}
\quad
& \frac{1}{T}
\sum_{t=1}^{T}
D_t(\mathbf{z}_t) \\
\mathrm{s.t.}\quad
& \tau B\log_2
\left(
1+\frac{|h_t|^2p_t}{\sigma^2}
\right)
\geq C_t(\mathbf{z}_t), \quad \forall t,\\
& \frac{1}{T}\sum_{t=1}^{T}p_t \leq \bar{P},\\
& 0\leq p_t \leq P_{\max}, \quad \forall t,\\
& z_{t,k}\in\{0,1\}, \quad \forall t,k,
\end{align}
\end{subequations}
where $\bar{P}$ is the average power budget and $P_{\max}$ is the peak power constraint.

Problem $\mathbf{P0}$ is a mixed-integer nonlinear programming problem. The binary variables determine the transmitted content, while the continuous variables determine physical-layer power allocation. Solving $\mathbf{P0}$ exactly is computationally expensive when $T$ and $K$ are large.

\subsection{Equivalent Power Cost Derivation}

For any fixed tile selection $\mathbf{z}_t$, the minimum power required to satisfy the rate constraint is obtained when the rate constraint is active:
\begin{equation}
    \tau B\log_2
    \left(
    1+\frac{|h_t|^2p_t}{\sigma^2}
    \right)
    =
    C_t(\mathbf{z}_t).
\end{equation}
Solving for $p_t$ gives
\begin{equation}
    p_t^{\min}(\mathbf{z}_t)
    =
    \frac{\sigma^2}{|h_t|^2}
    \left(
    2^{\frac{C_t(\mathbf{z}_t)}{\tau B}}-1
    \right).
\end{equation}
Therefore, the original problem can be equivalently transformed into a content selection problem:
\begin{subequations}
\begin{align}
\mathbf{P1}: \quad
\min_{\{z_{t,k}\}}
\quad
& \frac{1}{T}
\sum_{t=1}^{T}
\sum_{k=1}^{K}
a_{t,k}
\left[
(1-z_{t,k})d^{\mathrm{GS}}_{t,k}
+
z_{t,k}d^{\mathrm{corr}}_{t,k}
\right]\\
\mathrm{s.t.}\quad
& \frac{1}{T}
\sum_{t=1}^{T}
p_t^{\min}(\mathbf{z}_t)
\leq \bar{P},\\
& p_t^{\min}(\mathbf{z}_t)\leq P_{\max}, \quad \forall t,\\
& z_{t,k}\in\{0,1\}, \quad \forall t,k.
\end{align}
\end{subequations}
This transformation shows that the communication cost of transmitting a tile depends on both its bit size and the instantaneous wireless channel.

\begin{algorithm}[t]
\caption{GeoFovea-GS Scheduling Algorithm}
\label{alg:geofovea_gs}
\begin{algorithmic}[1]
\REQUIRE $\{d^{\mathrm{GS}}_{t,k}\}$, $\{d^{\mathrm{corr}}_{t,k}\}$, $\{a_{t,k}\}$, $\{C_{t,k}\}$, $\{|h_t|^2\}$, $\bar{P}$, $P_{\max}$.
\ENSURE $\{z_{t,k}^*\}$ and $\{p_t^*\}$.
\STATE Initialize $z_{t,k}=0$ and $\mathcal{S}_t=\emptyset$, $\forall t,k$.
\STATE Initialize the candidate set $\mathcal{A}=\{(t,k):t\in\mathcal{T},k\in\mathcal{K}\}$.
\STATE Compute $p_t=p_t^{\min}(\mathcal{S}_t)$, $\forall t$, where
\[
p_t^{\min}(\mathcal{S}_t)
=
\frac{\sigma^2}{|h_t|^2}
\left(
2^{\frac{C^{\mathrm{pose}}+\sum_{j\in\mathcal{S}_t}C_{t,j}}{\tau B}}
-1
\right).
\]
\IF{$\exists t$ such that $p_t>P_{\max}$ or $\frac{1}{T}\sum_{t=1}^{T}p_t>\bar{P}$}
    \STATE \textbf{return} infeasible.
\ENDIF
\WHILE{$\mathcal{A}\neq\emptyset$}
    \STATE Initialize feasible candidate set $\mathcal{F}=\emptyset$.
    \FOR{each candidate tile $(t,k)\in\mathcal{A}$}
        \STATE $\tilde p_{t,k}=p_t^{\min}(\mathcal{S}_t\cup\{k\})$.
        \STATE $\bar p_{t,k}^{\mathrm{new}}
        =
        \frac{1}{T}
        \left(
        \sum_{\ell=1}^{T}p_\ell+\tilde p_{t,k}-p_t
        \right)$.
        \STATE $\Delta D_{t,k}
        =
        a_{t,k}
        \max\{d^{\mathrm{GS}}_{t,k}-d^{\mathrm{corr}}_{t,k},0\}$.
        \STATE $\Delta p_{t,k}=\tilde p_{t,k}-p_t$.
        \IF{$\tilde p_{t,k}\leq P_{\max}$ and $\bar p_{t,k}^{\mathrm{new}}\leq\bar P$ and $\Delta D_{t,k}>0$}
            \STATE $\eta_{t,k}=\Delta D_{t,k}/(\Delta p_{t,k}+\epsilon)$.
            \STATE $\mathcal{F}\leftarrow\mathcal{F}\cup\{(t,k)\}$.
        \ENDIF
    \ENDFOR
    \IF{$\mathcal{F}=\emptyset$}
        \STATE \textbf{break}.
    \ENDIF
    \STATE Select $(t^\star,k^\star)=\arg\max_{(t,k)\in\mathcal{F}}\eta_{t,k}$.
    \STATE $z_{t^\star,k^\star}=1$.
    \STATE $\mathcal{S}_{t^\star}\leftarrow\mathcal{S}_{t^\star}\cup\{k^\star\}$.
    \STATE $p_{t^\star}\leftarrow p_{t^\star}^{\min}(\mathcal{S}_{t^\star})$.
    \STATE $\mathcal{A}\leftarrow\mathcal{F}\setminus\{(t^\star,k^\star)\}$.
\ENDWHILE
\STATE $z_{t,k}^*=z_{t,k}$ and $p_t^*=p_t$, $\forall t,k$.
\end{algorithmic}
\end{algorithm}

\begin{figure*} [t]
	\centering 
		 \includegraphics[width=1.0\textwidth]{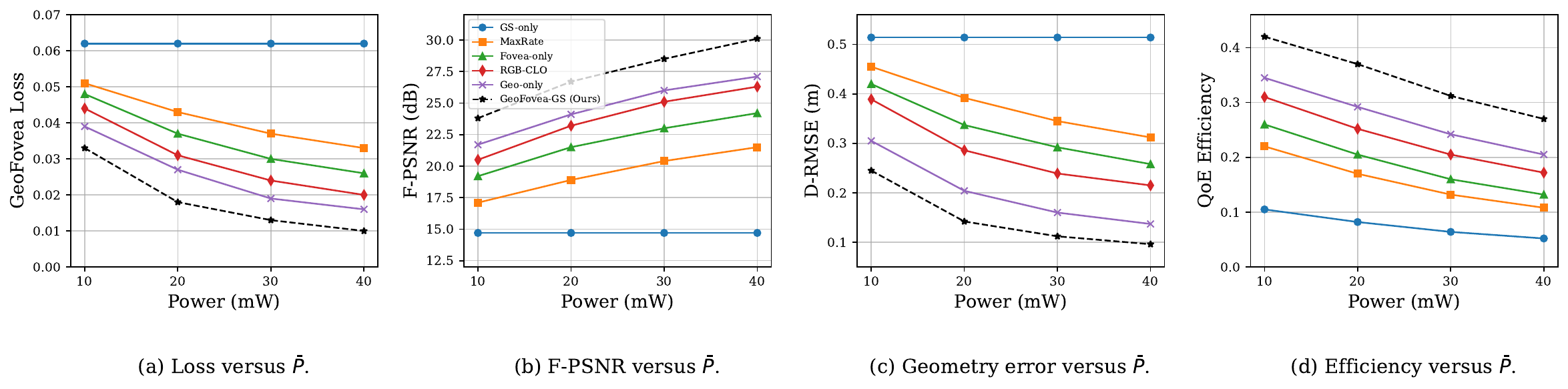}  
	
 	\caption{Quantitative comparison under different average power budgets. GeoFovea-GS consistently improves foveated rendering quality, geometry consistency,
and QoE efficiency.}  \vspace{-0.2in}\label{Fig3}
\end{figure*}

\subsection{Value-of-Information Scheduling}

To obtain a low-complexity solution, the distortion reduction obtained by transmitting tile $k$ at time slot $t$ is defined as
\begin{equation}
\Delta D_{t,k}
=
a_{t,k}
\max\left\{
d^{\mathrm{GS}}_{t,k}-d^{\mathrm{corr}}_{t,k},
0
\right\}.
\end{equation}
This is the amount of foveated geometry-aware distortion reduced when
$z_{t,k}$ changes from $0$ to $1$. In the case of high-quality correction
transmission, we set $d^{\mathrm{corr}}_{t,k}=0$.

Given the current selected set $\mathcal S_t$, the incremental power cost of transmitting tile $k$ is given by
\begin{equation}
    \Delta p_{t,k}
    =
    p_t^{\min}(\mathcal{S}_t\cup\{k\})
    -
    p_t^{\min}(\mathcal{S}_t),
\end{equation}
where $\mathcal{S}_t=\{k:z_{t,k}=1\}$ is the set of transmitted tiles at time slot $t$. The value-of-information score is then defined as
\begin{equation}
    \eta_{t,k}
    =
    \frac{\Delta D_{t,k}}{\Delta p_{t,k}+\epsilon},
\end{equation}
where $\epsilon>0$ is a small constant to avoid division by zero.

The score $\eta_{t,k}$ measures the VR distortion reduction per unit transmit power. A tile with large geometry error, high foveated importance, and low wireless cost receives a high priority.

\subsection{Complexity Analysis}

The proposed algorithm avoids exhaustive search over all tile selection patterns. Since there are $TK$ candidate tile transmissions, exhaustive search has complexity $\mathcal{O}(2^{TK})$. In contrast, Algorithm~\ref{alg:geofovea_gs} selects at most $TK$ tiles. If all scores are recomputed after each selection, the complexity is $\mathcal{O}(T^2K^2)$. With priority-queue implementation, the complexity can be reduced to approximately $\mathcal{O}(TK\log(TK))$.

The algorithm is therefore suitable for practical aerial VR systems, where the scheduler needs to make fast content adaptation decisions under time-varying wireless channels.

\section{Experiments}
In this section, the proposed GeoFovea-GS framework is evaluated on real-world UAV scenes. The experiments focus on three questions: 1) whether geometry-aware scheduling improves immersive rendering quality, 2) whether foveated and channel-aware resource allocation reduces wireless cost, and 3) whether the proposed method preserves fine structures and depth consistency in aerial VR views.

\begin{table}[t]
\centering
\caption{Quantitative comparison under $\bar{P}=20$ mW. Purple denotes the best value in each metric.}
\label{tab:main_quality}
\small
\setlength{\tabcolsep}{3.5pt}
\renewcommand{\arraystretch}{1.05}
\begin{tabular}{l|c c c c c}
\toprule
Method  & F-PSNR$\uparrow$ & F-LPIPS$\downarrow$ & D-RMSE$\downarrow$ & Flicker$\downarrow$ & Bits/F.$\downarrow$ \\
\midrule
GS-only       & 14.73 & 0.286 & 0.514 & 0.092 & \cellcolor{Best}0.0002 \\
MaxRate       & 18.92 & 0.219 & 0.392 & 0.073 & 0.410 \\
Fovea-only    & 21.46 & 0.172 & 0.337 & 0.061 & 0.408 \\
RGB-CLO       & 23.18 & 0.148 & 0.286 & 0.052 & 0.401 \\
Geo-only       & 24.05 & 0.132 & 0.204 & 0.047 & 0.392 \\
\midrule
GeoFovea-GS  & \cellcolor{Best}26.72 & \cellcolor{Best}0.094 & \cellcolor{Best}0.142 & \cellcolor{Best}0.035 & 0.371 \\
\bottomrule
\end{tabular}
\end{table}

\subsection{Baselines and Metrics}

GeoFovea-GS is compared with six baselines. \textbf{GS-only} transmits only the
camera pose and renders all views from the pre-trained 3DGS model.
\textbf{MaxRate} prioritizes tiles according to instantaneous channel quality.
\textbf{Fovea-only} selects tiles based only on gaze-centered foveated weights.
\textbf{RGB-CLO} uses RGB reconstruction error as the cross-layer content
importance score. \textbf{Geo-only} uses geometry inconsistency without
foveated weighting. \textbf{Full Image} transmits the full compressed image and
is used as a high-quality but communication-expensive reference in the
resource-cost comparison. 

To examine the quality of the scheduling solution, we also compare it with
\textbf{Exhaustive Search} on a reduced scheduling instance. Specifically, we
use $T_{\rm s}=3$ time slots and retain the top $K_{\rm s}=4$ candidate tiles
per slot according to their geometry-aware distortion, resulting in
$2^{T_{\rm s}K_{\rm s}}=4096$ tile-selection patterns. This setting is small
enough for exact enumeration while still preserving the most correction-relevant
tiles.

Visual quality is evaluated by F-PSNR, F-LPIPS, D-RMSE, and flicker. F-PSNR
and F-LPIPS are the foveated versions of PSNR and learned
perceptual image patch similarity (LPIPS), where pixel or feature errors are weighted according to the gaze-dependent importance map. D-RMSE denotes the root mean square error (RMSE) between the rendered depth and the reference geometry cue, and
flicker measures temporal instability across consecutive displayed views. Communication efficiency is evaluated by Bits/F., GeoFovea loss, and QoE efficiency, which correspond to transmitted bits per frame, the proposed foveated geometry-aware distortion, and quality gain per unit energy, respectively.

\begin{table}[t]
\centering
\caption{Resource cost to satisfy $\mathcal{L}_{\rm VR}\leq 0.02$. Purple denotes the best feasible result.}
\label{tab:resource_cost}
\small
\setlength{\tabcolsep}{4pt}
\renewcommand{\arraystretch}{1.05}
\begin{tabular}{l c|c c c c}
\toprule
Method & Feas. & Power$\downarrow$ & Bits/F.$\downarrow$ & Corr. Ratio$\downarrow$ & Violation$\downarrow$ \\
 & & (mW) & (Mbits) & (\%) & (\%) \\
\midrule
Full Image & $\checkmark$ & 52.4 & 5.42 & 100.0 & 14.8 \\
MaxRate    & $\times$     & --   & --   & --    & --   \\
Fovea-only & $\times$     & --   & --   & --    & --   \\
RGB-CLO    & $\checkmark$ & 24.7 & 0.61 & 11.2  & 3.9  \\
Geo-only   & $\checkmark$ & 20.6 & 0.48 & 8.5   & 2.1  \\
\midrule
GeoFovea-GS & $\checkmark$ & \cellcolor{Best}16.9 & \cellcolor{Best}0.37 & \cellcolor{Best}6.4 & \cellcolor{Best}0.8 \\
\bottomrule
\end{tabular}
\end{table}

\begin{figure} [t]
	\centering 
		 \includegraphics[width=0.48\textwidth]{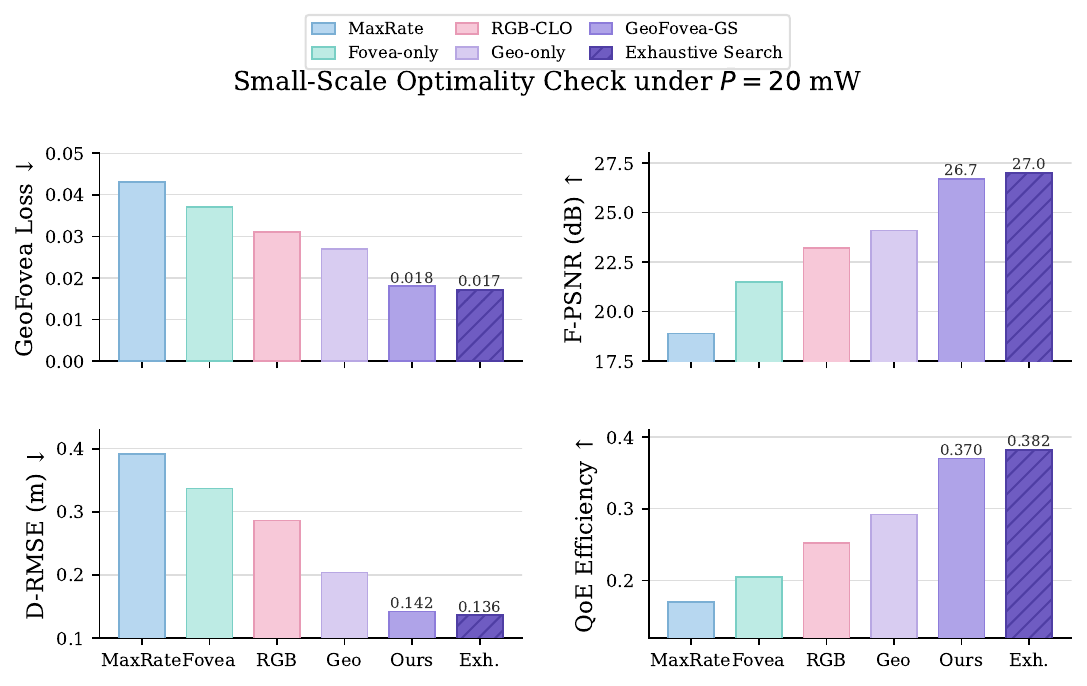} 
  	\caption{Small-scale optimality check under $\bar{P}=20$ mW. Exhaustive Search enumerates all feasible tile-selection patterns on reduced instances and serves as the optimum reference. GeoFovea-GS closely approaches Exhaustive Search across four metrics and consistently outperforms other scheduling baselines.}\vspace{-0.2in}  \label{Fig4}
\end{figure}

\begin{figure*} [t]
	\centering 
		 \includegraphics[width=1.0\textwidth]{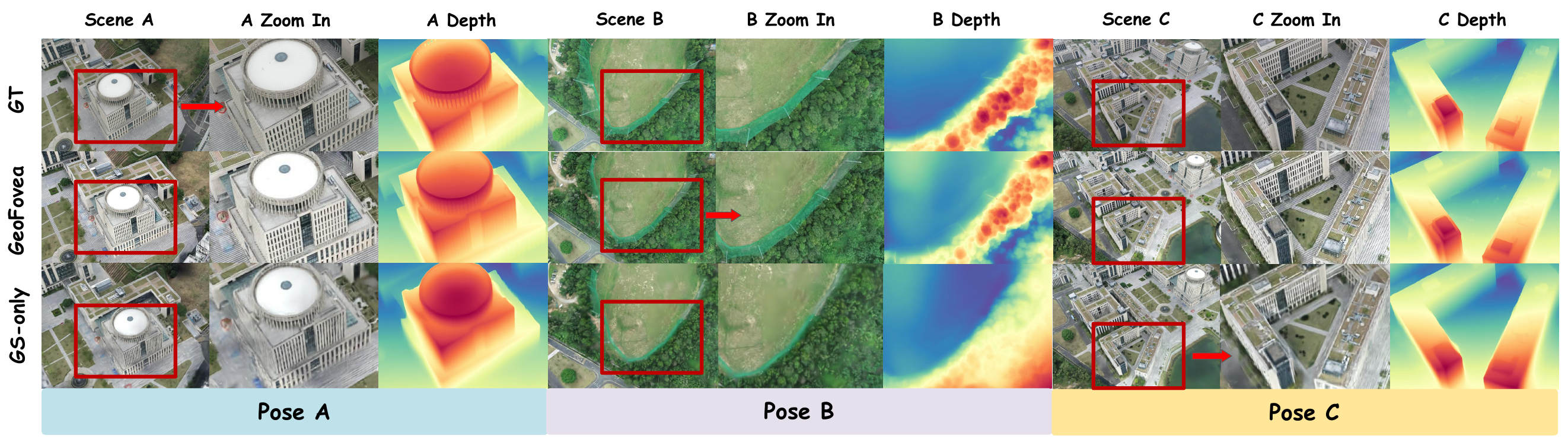}  
	
 	\caption{Visualization of wireless aerial VR reconstruction on SMBU. Each pose includes the rendered view, zoomed-in region, and depth map. Red boxes highlight geometry-sensitive regions. GeoFovea-GS preserves sharper visual structures and more consistent depth than GS-only.}\vspace{-0.2in} \label{Fig5}
\end{figure*}

\subsection{Dataset and Implementation}

GeoFovea-GS is evaluated on the SMBU scene from GauU-Scene
V2~\cite{xiong2024gauuscenev2assessingreliability}, which contains
large-scale UAV-captured outdoor views with buildings, roads, vegetation, and
open areas. The views are split by camera poses: training views are used to
construct the 3DGS scene memory, while held-out views with noticeable spatial
and angular displacement serve as VR query views and references. The 3DGS scene is trained on a single NVIDIA A100 GPU.

For correction transmission, the held-out reference tile is adopted as an
oracle high-quality correction, and $C_{t,k}$ is computed from its compressed
size. This protocol isolates the scheduling problem from the design of a
particular correction generator and evaluates the upper-bound gain achievable
when accurate scene-side correction is available. In a practical system, the
correction tile may be generated from nearby captured key views, multi-view
interpolation, or a high-quality offline renderer; such non-oracle correction
may reduce the absolute gain, but the proposed scheduler remains applicable
as long as an estimated correction distortion and bit cost are available.
The geometry cue is obtained from available scene depth and projected onto the
query image plane to compute tile-level geometry inconsistency. 

Each test view
is divided into non-overlapping tiles, and each tile is either rendered by
pose-only 3DGS or corrected through wireless transmission. Unless otherwise
specified, the wireless parameters follow Section~III, with $\bar P$ varied
from $10$ mW to $40$ mW. All results are averaged over testing views and random
channel realizations.

\subsection{Quantitative Comparison}

Table~\ref{tab:main_quality} compares the rendering quality and communication
cost under $\bar P=20$ mW. GS-only incurs negligible transmission cost but
suffers from severe photometric, geometric, and temporal artifacts, showing that
pose-only 3DGS rendering is unreliable for aerial VR. The adaptive baselines
improve over GS-only but address only part of the problem: MaxRate exploits
channel quality, Fovea-only emphasizes gaze importance, RGB-CLO relies on
photometric distortion, and Geo-only uses geometry reliability without
perceptual weighting. By jointly accounting for geometry reliability, foveated
importance, and wireless cost, GeoFovea-GS achieves the best F-PSNR, F-LPIPS,
D-RMSE, and flicker with only $0.371$ Mbits per frame. These results indicate
that wireless correction is most effective when resources are allocated to
regions that are simultaneously geometry-critical and perceptually important.

Fig.~\ref{Fig3} presents the performance under different average power budgets.
As $\bar P$ increases, the adaptive schemes achieve lower distortion and better
reconstruction quality, whereas GS-only remains unchanged because no correction
tiles are transmitted. GeoFovea-GS consistently yields the lowest GeoFovea loss
and the highest F-PSNR, especially in the low-power regime where only a small
number of tiles can be corrected. The clear D-RMSE gain over RGB-CLO further
shows that pixel-level distortion alone is insufficient for geometry-sensitive
aerial VR. The highest QoE efficiency also confirms that the selected
corrections provide larger quality gain per unit energy.

Table~\ref{tab:resource_cost} reports the resource cost required to satisfy
$\mathcal L_{\rm VR}\leq0.02$. Full Image transmission reaches the target but
requires prohibitive power and bandwidth, while MaxRate and Fovea-only are
infeasible because geometry reliability is not considered. RGB-CLO and Geo-only
meet the target at higher cost. GeoFovea-GS achieves the lowest power, bit rate,
correction ratio, and violation rate, requiring only $16.9$ mW and $0.37$ Mbits
per frame. This confirms that geometry-aware foveated scheduling can meet the
VR quality target with substantially fewer wireless resources

To further evaluate the quality of the proposed value-of-information scheduler, we conduct a small-scale optimality check where exhaustive search is computationally tractable. Specifically, we retain a reduced set of time slots and candidate tiles, enumerate all feasible tile-selection patterns, and select the solution with the minimum GeoFovea loss under the same wireless constraints. As shown in Fig.~\ref{Fig4}, GeoFovea-GS achieves performance close to Exhaustive Search across the four metrics. The small gap indicates that the proposed greedy scheduler captures most of the achievable distortion reduction, while avoiding the exponential complexity of exhaustive search. In contrast, MaxRate, Fovea-only, RGB-CLO, and Geo-only remain clearly inferior, confirming that jointly considering geometry reliability, foveated importance, and wireless power cost is essential even in the reduced problem.

\subsection{Visualization Results}

Fig.~\ref{Fig5} visualizes three representative testing poses from SMBU, each
showing the RGB view, a zoomed-in region, and the corresponding depth map. The
red boxes highlight geometry-sensitive regions where pose-only 3DGS rendering
produces blurred structures, distorted building boundaries, or unreliable depth,
especially around the circular rooftop in Pose A and the vegetation boundary in
Pose B. GeoFovea-GS corrects these high-impact regions, yielding sharper visual
details and more consistent depth transitions. This confirms that the proposed
scheduler improves not only pixel appearance but also geometry consistency in
perceptually important regions.

\section{Conclusion}
GeoFovea-GS was proposed as a geometry-aware crosslayer framework for communication-efficient wireless aerial
VR. By using 3D Gaussian Splatting (3DGS) as scene memory, reliable regions can be rendered from compact pose
information, while unreliable regions are corrected through
selective image/tile transmission. A foveated geometry-aware
distortion metric was developed to jointly capture photometric
error, depth inconsistency, and user-view importance. Based on
this metric, a lightweight value-of-information scheduler was
designed to allocate wireless resources to geometry-critical
and perceptually important tiles. Experiments on real-world UAV scenes show that, compared with Geo-only, the
strongest adaptive baseline in Table~I, GeoFovea-GS improves F-PSNR by
$11.1\%$ and reduces depth error by $30.4\%$. Under the target
$L_{\rm VR}\leq 0.02$, it further reduces the required power and bit rate by
$18.0\%$ and $22.9\%$, respectively, relative to Geo-only, and reduces the bit
rate by $39.3\%$ relative to RGB-CLO.

\bibliographystyle{IEEEtran}
\bibliography{ref}

\end{document}